\documentclass[%
aip,
jcp,
10pt,
amsmath,
amssymb,
nofootinbib
]{revtex4-2}

\usepackage{titlesec}
\titlespacing*{\section}
{0pt}{1.5ex}{1.5ex}
\titlespacing*{\subsection}
{0pt}{1ex}{1ex}
\titlespacing*{\subsubsection}
{0pt}{1ex}{1ex}
\usepackage{setspace}
\usepackage{tabularx}
\usepackage{wasysym}

\usepackage[version=3]{mhchem} 
\usepackage[dvipsnames]{xcolor}
\usepackage{amsfonts}

\usepackage{graphicx}
\usepackage[colorlinks=true,citecolor=red]{hyperref}

\usepackage[dvipsnames]{xcolor}
\usepackage{url}
\usepackage{color}
\usepackage{dcolumn}
\usepackage{bm}
\usepackage[mathlines]{lineno}
\usepackage{comment}
\usepackage[version=3]{mhchem}

\usepackage[capitalise]{cleveref} 

\usepackage{graphicx} 
\usepackage{amsmath}
\usepackage{longtable}
\usepackage{wrapfig,floatflt}
\usepackage{etoolbox}
\usepackage{times}
\usepackage{siunitx}

\usepackage{multirow}
\renewcommand{\bm}[1]{\mathbf{#1}}
\newcommand*{\myfont}{\fontfamily{phv}\selectfont}
\DeclareTextFontCommand{\textmyfont}{\myfont}

\newcommand*\jy[1]{\textcolor{black}{#1}}

\let\oldbibliography\thebibliography
\renewcommand{\thebibliography}[1]{\oldbibliography{#1}
\setlength{\itemsep}{0pt}} 
\allowdisplaybreaks

\makeatletter
\renewcommand{\thesection}{\arabic{section}}
\renewcommand{\thesubsection}{\arabic{section}.\arabic{subsection}}
\renewcommand{\p@subsection}{}

\renewcommand{\p@subsubsection}{}
\makeatother

\begin{document}


\title{Temperature-Dependent Ion Dynamics in Argyrodite Superionic Conductor Li$_{6}$PS$_{5}$Cl with Fine-Tuned MACE Foundation Model}

\author{Jack Yang}
\affiliation{School of Materials Science and Engineering, University of New South Wales, Kensington, New South Wales 2052, Australia}
\email{jianliang.yang1@unsw.edu.au}

\author{Lei Ao}
\affiliation{Jiangxi Provincial Key Laboratory of Advanced Electronic Materials and Devices, Jiangxi Science and Technology Normal University, Nanchang 330018, China}

\author{Sean Li}
\affiliation{School of Materials Science and Engineering, University of New South Wales, Kensington, New South Wales 2052, Australia}

\begin{abstract}

The rapid growth of machine-learning interatomic potentials (MLIP), particularly  the foundational models for materials chemistry, such as MACE, has significantly strengthened  our capabilities in studying the complex dynamics of materials through statistical sampling. Here, by performing molecular dynamics with a fine-tuned MACE model, that are further combined with essential dynamics analysis, a tool that is commonly applied in (bio)molecular simulations, we examine the temperature-dependent ion dynamics in a prototypical inorganic solid superionic conductor, namely argyrodite \ce{Li6PS5Cl}. Our analyses show clear localised phonon-like dynamics for \ce{Li+} ions below 500 K with trajectories lying in a low-dimensional manifold that is highly nonlinear, which is opposite to the case in the high-temperature ($\sim$800 K) end. Between 500-700 K, the system exhibit strong competition in the dynamics between the \ce{Li+} and framework ions, that are reflected in both the eigenspectrum and the projected low-dimensional embeddings characterising these dynamics not showing a simple monotonic dependency upon rising temperatures, and is further supported by investigations employing Langevin dynamics that systematically examine how damping the framework motions affects the dynamics of the \ce{Li+} ions, thanks to the computational efficiency provided by MLIP. This work thus contributes towards the broader theoretical and data-driven framework for studying the lattice dynamics of solids.     

\end{abstract}	
\maketitle

\section{Introduction}

Superionicity\cite{mahan1976superionic} represents a unique state of solid that lies in the intersection between an ordered crystal and a random liquid. Below the transition temperatures, all ions in the materials exhibit localised vibrations that confine with the standard phonon model for lattice vibrations,\cite{born1996dynamical} whereas above the transition temperature, the mobile ions enter their superionic state, which diffuse across long distances through the crystal like a liquid, whilst the framework itself retains a phonon-like vibrating state. Such  unique ionic dynamics have attracted considerable research interest over decades\cite{parrinello1983structural,jun2024diffusion} for their potential applications as solid-state batteries\cite{kwak2022emerging} and thermoelectrics,\cite{zhao2020recent}  with exemplary materials including $\alpha$-\ce{AgI},\cite{tatsumisago1991stabilization} \ce{Na3PS4},\cite{gupta2021fast} \ce{Li3YCl6}\cite{ahammed2024configurational} and \ce{Li6PS5Cl}.\cite{ding2025liquid}

Nevertheless, the ionic dynamics in superionics are still yet to be fully understood from a theoretical perspective, as the harmonic phonon model from standard condensed matter physics is no longer fully adequate to describe the large-amplitude liquid-like ionic diffusions, even with the inclusion of higher-order anharmonic terms.\cite{cowley1968anharmonic} This renders the use of long-time molecular dynamics (MD) simulations with alterative theoretical frameworks to analyse, understand, and extract useful information on the ion dynamics in superionics. This inspires the present investigation, using the argyrodite \ce{Li6PS5Cl} as an exemplary material system that is based on the recently published inelastic neutron scattering investigations on the liquid-like \ce{Li+} ion dynamics by \citet{ding2025liquid}, demonstrating an excessive Li phonon density-of-states (PDOS) at the long wavelength limit that vary linearly with respect to the vibration frequency, as opposed to the quadratic proportionality in the standard Debye model. 

The present investigation essentially stands on two theoretical basis.

Firstly, the recent advancement in foundational energy models,\cite{yuan2026foundation} in particular, the MACE (Multi-Atomic Cluster Expansion)\cite{batatia2022mace} suite of models,\cite{batatia2023foundation} that present the community with a broadly applicable machine-learning interatomic potentials (MLIPs) at the accuracy of DFT, while overcoming its $\mathcal{O}(N^3)$ scaling bottleneck, making them comparable to empirical atom-atom forcefields in terms of the computational speed (with modern GPU accelerations), to enable nanosecond-scale MD samplings with $\geq10^3$ atoms. Most importantly, extending the applicability of the pre-trained foundation models to new systems can generally be achieved with small dedicated training sets and significantly less number of epochs, a process so-called \emph{fine-tuning}, making it an attractive energy model for tackling a wide range of emerging and challenging material problems, such as the investigation of the dynamics of superionics.

Secondly, we adopted the framework of essential dynamic analysis,\cite{daidone2012essential} which is a method that has been more widely applied in (bio)molecular simulation, and conceptually similar to principal component analysis,\cite{jolliffe2016principal} to extract the dominating ion dynamics from MLIP-MD trajectories. This essentially provides a single mathematical framework that is applicable across both the localised phonon-like vibration and large-amplitude ionic diffusion regimes. It further enables us to construct visually clear projections onto low-dimensional manifold, from which we can apply differential geometric analysis to further unpick the  structures of the complex geometric space in which the collective dynamics of ions take place.

With these, we start from fine-tuning the MACE foundation model to reproduce the key vibrational characteristics of \ce{Li6PS5Cl}, with which, $NVE$ and Langevin MLIP-MD sampling across multiple temperatures, as well as damping strengths imposed on the framework dynamics, are conducted to systematically examine the correlated ionic  motions  in \ce{Li6PS5Cl}. Essential dynamics and geometric analysis that are performed on the MLIP-MD trajectories clearly reveal the low-temperature localised phonon-like behaviours of the \ce{Li+} ions, with the trajectories lying in a highly nonlinear low-dimensional manifold. This is in stark contrast to the geometric structures of the trajectory sampled in the high-temperature superionic regime, in which the trajectories are dominated by large-amplitude motions of the \ce{Li+} ion over a much smoother manifold. On the other hand, the transition temperature regime are characterised by the competition in the dynamics between the framework and \ce{Li+} ions, resulting in a complex temperature dependency and geometric structure that characterise the collective motion of the \ce{Li+} ions. Extending the current conceptual framework, which combine MLIP-MD, essential dynamics and geometric analysis, could further deepen our fundamental understanding on how superionic transport is facilitated by the framework dynamics, particularly those with strong anharmonicity,\cite{tatsumisago1991stabilization,gupta2021fast} in the condensed phase.  

\section{Computational Methodologies}
\label{sect:methodology}

\noindent \textit{\textbf{Ab initio Molecular Dynamics (AIMD)}}  The Vienna \textit{Ab initio} Simulation Package (VASP)\cite{kresse1996efficient}  with the standard projector augmented wave (PAW) method\cite{kresse1999ultrasoft} was employed to perform the AIMD simulations on \ce{Li6PS5Cl}. More specifically, the structure \ce{Li6PS5Cl} was retrieved from the Materials Project\cite{jain2013commentary} database where its unit cell and ionic position were first optimised with the PBE\cite{perdew1996generalized} functional. The structure was then equilibrated from 10 to 710 K using the velocity-scaling algorithm for 8.75 ps, after which a production trajectory of 5 ps long was simulated under the $NVE$ ensemble, both at 1 fs step. All calculations were performed with a single $\Gamma$-centred K-point and a pseudopotential cutoff of 520 eV. 

\noindent \emph{\textbf{Foundation Model Fine-Tuning}} All AIMD frames from both the equilibration and production trajectories were used  for fine-tuning the MACE foundation energy model (\texttt{omat-0-medium}),\cite{batatia2023foundation} with 5\% of the data reserved for validation. 
The pre-determined weights on the energy, force and stress loss functions were 1, 10, and 100, respectively to construct the Huber loss function for training. The learning rate was set at $5\times10^{-4}$.
In total, 400 epochs\footnote{Practically, $\sim 20$ epochs would be sufficient for fine-tuning the foundation model to reach same level of accuracy reported here, as we did not observe significant improvement in the model performance with more fine-tuning epochs.} of fine-tuning were performed with the final model reaching $5.903\times 10^{-2}$ meV/atom in the root-mean-squared-error (RMSE) for energy, 11.30 meV/$\mbox{\AA}$  for atomic forces and $6.860\times10^{-2}$ meV/$\mbox{A}^3$ for stress.  

\noindent \emph{\textbf{Molecular Dynamic Simulations with MLIP}} The Atomic Simulation Environment (ASE)\cite{larsen2017atomic} Python package was deployed to perform the MD simulations on \ce{Li6PS5Cl} with  MLIP (which is the fine-tuned MACE energy model discussed above). The simulations were carried out using a $(3\times3\times3)$ supercell containing 1,404 atoms in total. The system was first equilibrated to the target temperature under the $NPT$ ensemble for 50 ps, followed by the production run under the $NVE$ ensemble  for 1 ns. All MD simulations were performed at 1 fs step size, and snapshots were collected every 10 fs for subsequent analysis.

\noindent \emph{\textbf{Phonon Density-of-States}} We extract the finite-temperature PDOS from the Fourier transformation of the velocity autocorrelation function (VACF) that is computed from the MLIP-MD simulations at a given temperature. The VACF for the $n-$th particle is defined based on the its velocity as 
\begin{equation*}
    \label{eq:vacf}
    C_n(t)=\langle\bm{v}_n(0)\bm{v}_n(t)\rangle_t=\lim_{\tau\to\infty}\int_0^\tau \bm{v}_{n}(t^\prime)\bm{v}_{n}(t^\prime+t)dt^\prime,
\end{equation*}
which, upon Fourier transformation, leads to the frequency-dependent PDOS as 
\begin{equation}
    \label{eq:PDOS_FT}
    g(\omega)=\sum_{n=1}^N\int C_{n}(t)e^{i\omega t}dt.
\end{equation}
In order to extract the elemental contributions to the total PDOS, the following normalisation scheme is used. For all element types $a\in\mathcal{A}$, we compute $g_a(\omega)$ by restricting the summation in \cref{eq:PDOS_FT} to all the atoms $N_a$ with element type $a$ in the host lattice. The total PDOS is then given by $g_{\mathrm{total}}(\omega)=\sum_a g_a(\omega)$. For this additive formalism to work, (a) we do not pre-normalise $C_{n}(t)$ by its initial value of $C(0)$, and (b) we then enforce the normalisation condition of $\int_0^{\omega_{\max}}g_{\mathrm{total}}(\omega)d\omega=1$, by renormalising the raw values computed for both $g_{a}(\omega)$ and $g_{\mathrm{total}}(\omega)$ with $1/\int_0^{\omega_{\max}}g_{\mathrm{total}}(\omega)d\omega$. 

\section{Results and Discussions}

\subsection{Phonon density-of-states from the foundation model}

We begin our discussions by first examining the capability of the MACE foundation model in reproducing the key phonon features as reported in \citet{ding2025liquid} through computing the VACF from the MLIP-MD trajectories. This not only allow us to take into account the effects of thermal-induced phonon frequency-shift and broadening, but more importantly, the diffusive behaviour of Li atoms, which is otherwise ignored in the conventional phonon calculations based on the finite-displacement approach.\cite{togo2015first}

The baseline foundation model chosen is the \texttt{omat-0-medium}\cite{batatia2023foundation} variant, which was shown to be the best performing model for reproducing the lattice dynamics of halide double perovskites from our previous benchmark study.\cite{yang2026mace} However, our initial trial using the  pre-trained \texttt{omat-0-medium} foundation model  showed that the computed PDOS only spanned approximately half of the frequency range, compared to the reported PDOS for \ce{Li6PS5Cl} in \citet{ding2025liquid}, and such a finding is consistent with the benchmark results reported by \citet{deng2025systematic} showing the systematic softening of the MLIPs for inorganic solids. As with most MLIP,\cite{nigam2022unified} MACE\cite{batatia2022mace} surrogates the potential energy surfaces predominantly from learning the short-ranged local atomic environments, the highly mobile Li ion in superionic conductors, together with the unique structures of \ce{Li6PS5Cl} that enables, for example, the rattling motion of the PS$_4$ tetrahedra, can induce strong variations in the  local chemical environments at finite temperatures that can become out-of-distribution for the pre-trained foundation model. 

\begin{figure*}[tbh]
    \centering
    \includegraphics[width=\linewidth]{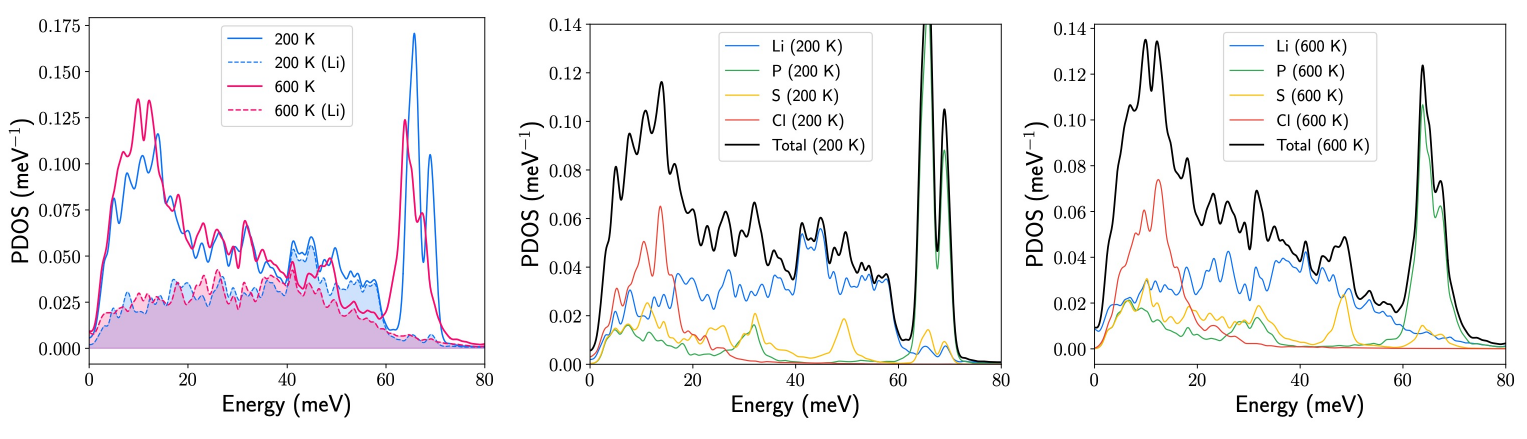}
    \caption{\textbf{Phononic properties of \ce{Li6PS5Cl} determined from MLIP-MD.} Finite temperature PDOS at 200 and 600K for \ce{Li6PS5Cl} computed from the Fourier transformation of VACF from NVE-MD trajectories sampled with the fine-tuned MACE foundation model. (left) Comparing the total and Li PDOS at 200 and 600 K. (middle) and (right) compared the total and element-resolved (Li, P, S and Cl) PDOS at 200 and 600 K, respectively.}
    \label{fig:LiPSCl_PDOS_MACE_FT}
\end{figure*}

This suggests that it is necessary to fine-tune the \texttt{omat-0-medium} model before proceeding for further analysis. As detailed in \cref{sect:methodology}, the data used for model fine-tuning consists of AIMD trajectory sampled from 10 to 700 K under the velocity-scaling scheme, as well as additional trajectory sampled at 700 K under the \emph{NVE} ensemble, whereby the inclusion of  MD structures across a wide temperature range, particularly the high-temperature ones for model training, has demonstrated its efficacy in enhancing the stabilities of long-time MD simulations using MLIP.\cite{fu2022forces,yang2025accounting} \cref{fig:LiPSCl_PDOS_MACE_FT} shows the total and element-resolved partial PDOS for \ce{Li6PS5Cl} computed from \emph{NVE} MD using the fine-tuned  \texttt{omat-0-medium} model, which reproduced some of the key features reported by \citet{ding2025liquid}, including: (a) Significant contribution from the Li ion vibration in the energy range of 0-60 meV, with higher PDOS for Li at 200 K in the energy range between 40-60 meV [\cref{fig:LiPSCl_PDOS_MACE_FT}(left)]. (b)  Dominant contribution from the Cl vibration below 20 meV as shown in \cref{fig:LiPSCl_PDOS_MACE_FT}(middle and right). (c) Dominant contribution from P vibration between 60-75 meV in the PDOS [\cref{fig:LiPSCl_PDOS_MACE_FT}(middle and right)] with a blue shift in the peak positions upon heating [\cref{fig:LiPSCl_PDOS_MACE_FT}(left)]. And most importantly, (d) a finite PDOS at 0 meV that is predominantly contributed by Li at 600 K [\cref{fig:LiPSCl_PDOS_MACE_FT}(right)] as one of the key signature for liquid-like Li ion conduction behaviour in  \ce{Li6PS5Cl}. 

However, as shown in \cref{fig:LiPSCl_PDOS_MACE_FT}(middle), our simulation also reveals a small non-zero PDOS at the zero frequency at 200 K with predominant contribution from Cl vibration, that is inconsistent with the previous result.\cite{ding2025liquid}  In order to mitigate the problem of finite PDOS at the zero frequency at 200 K, which tends to suggest that the low-temperature data may not be well represented in the training set, we also tried other strategies of model fine-tuning, including (a) further fine-tuning the above fine-tuned MLIP with extra AIMD trajectory sampled at 200 K, and (b) fine-tuning \texttt{omat-0-medium} using  the AIMD equilibration trajectory only. Nevertheless, we found that these two fine-tuning approaches  worsen the model performances such that the 200K-PDOS at the zero frequency was further increased. In this case, the currently observed finite PDOS may be a problem that is inherited from the nature of the data that is used for training the baseline \texttt{omat-0-medium} model, rendering ongoing research efforts to understand the interplay among the nature of the training data, model complexity, intrinsic accuracy of the model and its practical applicability in realistic simulations,\cite{creed2026six} that is beyond the scope of the present work.\footnote{It  only came to our attention at the later sate of this project that a more robust  protocol for training the MACE-based interatomic potentials, that is more specifically targeted at \ce{Li6PS5Cl} as the model system, has been published by \citet{fragapane2026li}.}  Since most of key features in PDOS for \ce{Li6PS5Cl} have nonetheless been reproduced with the fine-tuned model, we shall proceed our subsequent discussions based on the MD results that are generated with this fine-tuned MACE model.

\subsection{Essential dynamics analysis}
\label{sec:essential_dynamics}

To gain better understanding on the temperature-dependent ionic dynamics of \ce{Li6PS5Cl}, we perform the essential dynamics analysis\cite{daidone2012essential} on the MD trajectories. Fundamentally, essential dynamics analysis is built upon the concept of principal component analysis, which can facilitate the identification of large-amplitude collective atomic movements in the crystal. Drawing analogy to the Hessian for phonon calculations, the starting point here is to construct the two-body covariance matrix $\bm{C}$ for (mass-weighted) atomic displacements, with the matrix elements $C_{ij}^{\alpha\beta}=\langle\sqrt{m_i m_j}\Delta r_{i}^{\alpha} \Delta r_{j}^{\beta}\rangle_t$, where $i$, $j$ are the atomic indices and $\alpha$, $\beta$ denote one of the three orthogonal Cartesian directions, $\langle\cdot\rangle_t$ represents statistical averaging over time across the entire MD trajectory. The atomic displacement is computed with respect to its averaged atomic position as 
\begin{equation}
\label{eq:instant_dispalcement}
    \Delta\bm{r}_i(t)=\bm{r}_i(t)-\langle\bm{r}_{i}\rangle_t.
\end{equation}
that is corrected with the periodic boundary conditions. Unlike the Hessian matrix that only accounts for the local curvature of the potential energy surface under a harmonic approximation, the covariance matrix $\bm{C}$ naturally incorporate the correlations between long-ranged displacements in $\Delta \bm{r}$ directly by construction.

Solving the  eigenvalue problem of $\bm{C}\bm{v}_k=\lambda_k\bm{v}_k$, and sorting the eigenvectors $\{\bm{v}_k\}$ in the descending order of the corresponding eigenvalues $\{\lambda_k\}$ will provide us with information on the collective coordinates (\emph{c.f.} vibrational eigenmodes) that entail large amplitude collective movements of ions (with large $\lambda_k$) in the solid.  In particular, we can further partition the full covariance matrix into blocks that characterise the motions of Li-ions only $\bm{C}_{\mathrm{Li-Li}}$, framework-ions only $\bm{C}_{\mathrm{FM-FM}}$, as well as the coupled motions between the Li and framework ions $\bm{C}_{\mathrm{Li-FM}}$:
\begin{equation*}
    \bm{C}=\left[\begin{array}{cc}
       \bm{C}_{\mathrm{Li-Li}}  &  \bm{C}_{\mathrm{Li-FM}} \\
       \bm{C}_{\mathrm{Li-FM}}^\top  & \bm{C}_{\mathrm{FM-FM}}
    \end{array}\right].
\end{equation*}
Here, by analysing the eigenvalue spectrum for the full covariance matrix $\bm{C}$ and the Li-only sub-block $\bm{C}_{\mathrm{Li-Li}}$ will provide us with the critical insights on the interplay between the dynamics of the Li and framework ions in \ce{Li6PS5Cl} across the practically relevant range of temperatures.


\begin{figure*}[tbh]
\centering
\includegraphics[width=\linewidth]{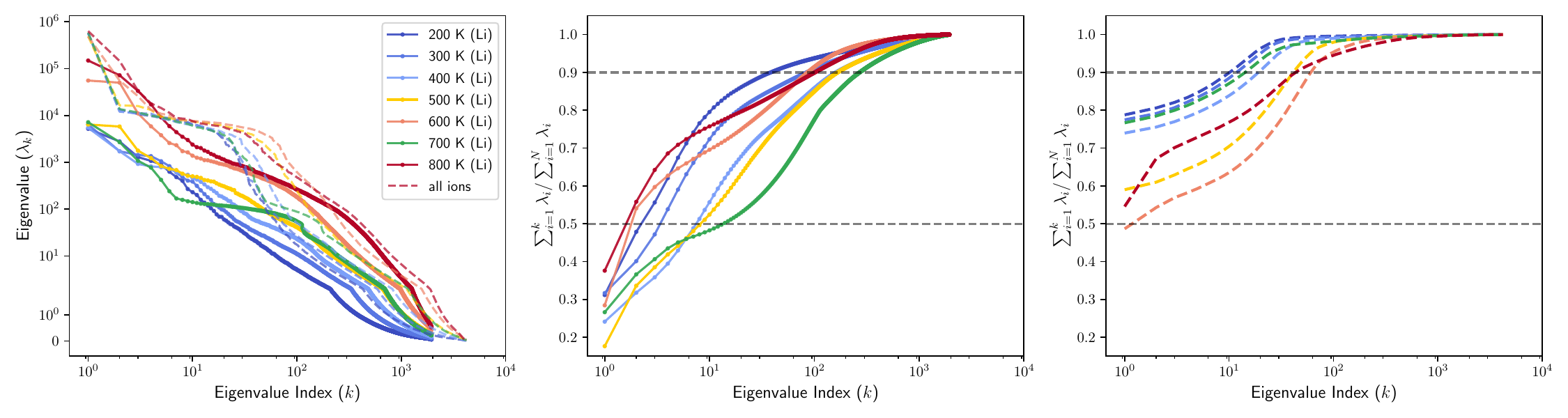}
\caption{\textbf{Temperature-dependent eigenvalue spectrum for essential dynamic analysis under $NVE$ ensemble.} (left) Sorted eigenvalues $\lambda_k$  of the covariance matrices characterising the Li-ion-only (solid curves with circled markers) and all-ion movements (dashed curves) computed based on the $NVE$ trajectories sampled at different temperatures. 
Participation ratios ($\sum_{i=1}^k \lambda_i / \sum_{i=1}^N \lambda_i$) for the Li-ion-only (middle) and all-ion (right) as a function of the eigenvalue index $k$, signifying the percentages of eigenvectors needed to describe the collective movements of ions in the system. Dashed horizontal lines correspond to 50 and 90 \% of ionic movements.}
\label{fig:nve_md_essential_dynamics_analysis}
\end{figure*}

\cref{fig:nve_md_essential_dynamics_analysis} compares the eigenvalue spectrums for $\bm{C}_{\mathrm{Li-Li}}$  and $\bm{C}$ that characterise how thermal energies activate the ionic dynamics in \ce{Li6PS5Cl}. In particular, the eigenvalues of $\bm{C}_{\mathrm{Li-Li}}$ characterise the dynamics of the Li ions under the influences from the fluctuating crystal potentials induced by the vibrating frameworks, which can be understood analogously from the concepts behind the temperature-dependent-effective-potential (TDEP)\cite{hellman2013temperature} for anharmonic phonons. 

\begin{figure*}[tbh]
    \centering
    \includegraphics[width=\linewidth]{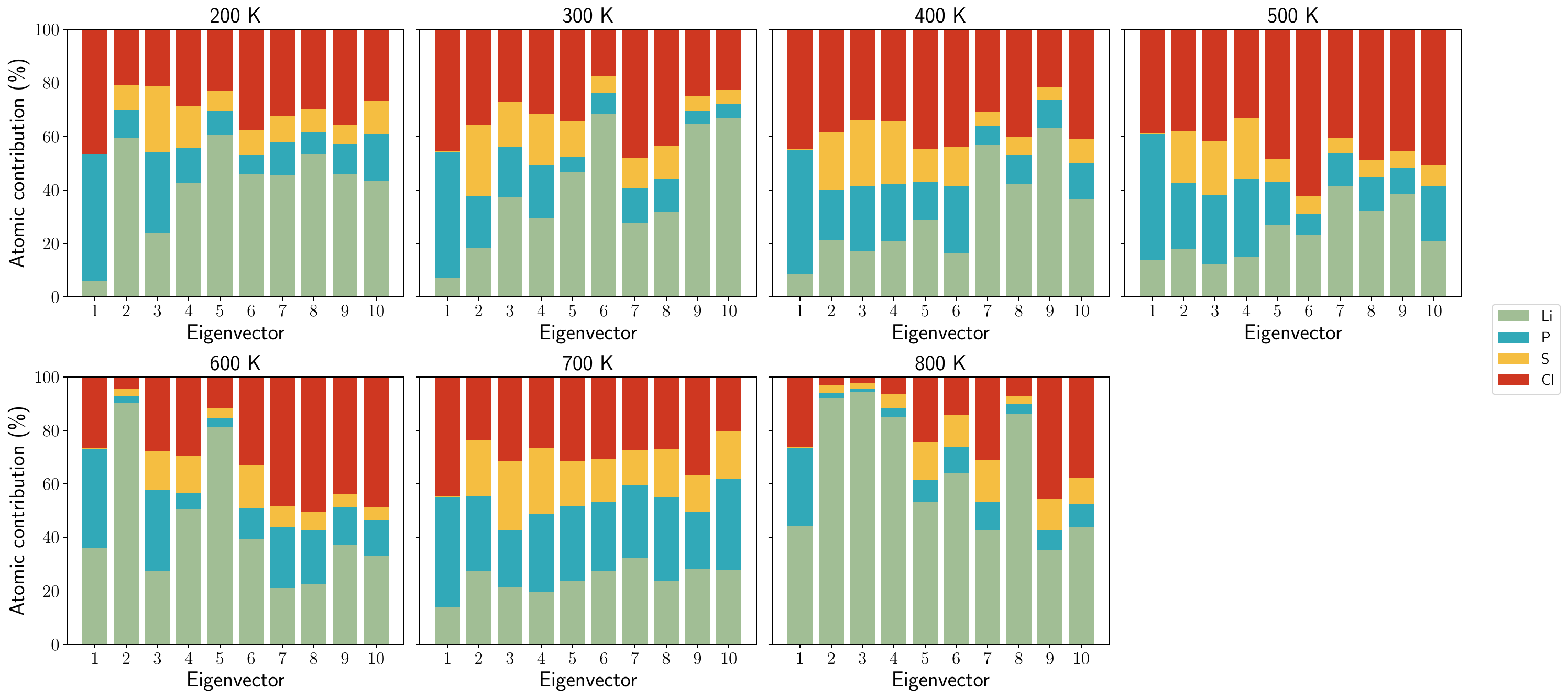}
    \caption{\textbf{Atomic contributions to the first 10 leading eigenvectors of the covariance matrix.} Participation from each elements in \ce{Li6PS5Cl}  to the first 10 leading eigenvectors of the covariance matrix $\bm{C}$ at different simulation temperatures. For each atom $i$ in the simulation supercell, we first compute the $\ell 2$-norm of the eigenvector on that atom from its Cartesian components $p_i^{(n)}=\sqrt{\left(\bm{e}_{ix}^{(n)}\right)^2+\left(\bm{e}_{iy}^{(n)}\right)^2+\left(\bm{e}_{iz}^{(n)}\right)^2}$ where $n$ is the index of the eigenvector, from which the participation ratio for a given type of element $\mathcal{A}$ is $p_a^{(n)}=\sum_{i\in\mathcal{A}}p_i^{(n)}/\sum_{i}p_i^{(n)}$. 
    }
\label{fig:nve_essential_dynamics_atom_contributions}
\end{figure*}

Focusing first on the $\lambda_k$'s for  $\bm{C}_{\mathrm{Li-Li}}$ [solid lines in \cref{fig:nve_md_essential_dynamics_analysis}(left)], we see a strong fluctuations in the distributions of $\{\lambda_k\}$ in  between 500 and 800 K, which crosses the critical temperature  ($\sim$ 600 K) at which the liquid-like ion conduction behaviours are expected to occur.\cite{ding2025liquid} More specifically, when temperature is raised from 500 to 600 K,  we see that the eigenvalues are significantly increased by an order of magnitude. This is also demonstrated in \cref{fig:nve_essential_dynamics_atom_contributions},  where the atomic contributions from  Li to the components in the leading \jy{five eigenvectors increased at 600 K, compared to the lower temperature cases}, which again reflects the \emph{liberation} of \ce{Li+} ions to undergo large amplitude movements at the \jy{elevated} temperature. Such an effect is further reflected when we analyse the accumulated mode participation ratios [\cref{fig:nve_md_essential_dynamics_analysis}(middle)] for $\bm{C}_{\mathrm{Li-Li}}$, whereby at 500 K, around 10 $\bm{v}_{k}$ are needed to describe 50 $\%$ of the collective  motions of the \ce{Li+} ions, but this is achievable just with the two leading eigenvectors at 600 K.

As the temperature is further raised to 700 K, the system enters another interesting dynamic regime, where the eigenvalues for $\bm{C}_{\mathrm{Li-Li}}$ are notably reduced again to the level that is comparable to the eigenvalues of $\bm{C}_{\mathrm{Li-Li}}$ computed  at 500 K [\cref{fig:nve_md_essential_dynamics_analysis}(left)], leading to a consequence in which the most number of eigenmodes are required to describe up to 50 \% of the collective motions of the \ce{Li+} ions  across all the temperatures considered here [\cref{fig:nve_md_essential_dynamics_analysis}(middle)]. Geometrically, it demonstrates that the diffusion pathway of the \ce{Li+} ions lies in the most complex manifold in terms of dimensionality at this particular temperature window.   This is a result of the fact that, at 700 K, the leading 10 eigenvectors for $\bm{C}$ exhibited the strongest contributions from the movements of framework ions [\cref{fig:nve_essential_dynamics_atom_contributions}], which is also why simply using the first eigenvector of $\bm{C}$ is sufficient to characterise over 50 \% of the ionic movements in the system [\cref{fig:nve_md_essential_dynamics_analysis}(right)] showing a stark contrast to the result shown in \cref{fig:nve_md_essential_dynamics_analysis}(middle). This suggests that, at 700 K, the thermal energy becomes sufficiently large to excite the large amplitude vibrations of the framework, which now moves concomitantly with the Li sublattice to facilitate the diffusion of \ce{Li+} ions. When the temperature rises to 800 K, the lighter \ce{Li+} ions gain sufficient extra thermal energy, making the ionic dynamics to be dominated by them again.

\jy{Further comparing the eigenvalues for $\bm{C}_{\mathrm{Li-Li}}$ and the full covariance matrix $\bm{C}$ (which shall be denoted as $\{\lambda_{\mathrm{Li-Li}}\}$ and $\{\lambda\}$, respectively), we can better observe  the signature on how the \ce{Li+} ion movements become more dominant at the higher temperatures. As shown in \cref{fig:nve_md_essential_dynamics_analysis}(left), $\{\lambda\}$ are generally larger than $\{\lambda_{\mathrm{Li-Li}}\}$ across the full temperature range being investigated, due to the fact that the former reflects the participation from all atoms in \ce{Li6PS5Cl} towards the collective atomic motions under thermal excitations, which is also observable from the \emph{per-element}  participation ratios in the leading eigenmodes of $\bm{C}$ (\cref{fig:nve_essential_dynamics_atom_contributions}). Consequentially, unlike $\bm{C}_{\mathrm{Li-Li}}$, the first leading eigenvector $\bm{v}_1$ for $\bm{C}$ is sufficient for describing more than 50 \% of the collective motion of the entire system [\cref{fig:nve_md_essential_dynamics_analysis}(right)]. Below 500 K, the participation ratio for $\bm{v}_1$  is even higher than 70 \%. However, the participation ratios of $\bm{v}_1$ at both 600 and 800 K are seemingly lower, due to the fact that the lattice dynamics become strongly dominated by that of the \ce{Li+} ions, resulting in the distributions of $\{\lambda_{\mathrm{Li-Li}}\}$ and $\{\lambda\}$ become very close to each other [\cref{fig:nve_md_essential_dynamics_analysis}(left)]. }

\subsection{Geometric signatures of the collective dynamics of the lithium ions} 

Beyond identifying the dominant collective motions from the eigen-spectrums of the essential dynamics matrices, we can proceed further to characterise the geometric manifold, that corresponds to the configurational space in which the \ce{Li+} ions reside. This will provide us with further insight into how the underlying ionic dynamics evolve with temperature. 

From a differential-geometric perspective, the configurations of the \ce{Li+} ions in the crystal framework of  \ce{Li6PS5Cl} that have been sampled in MD simulations  can be viewed as a point cloud in a high-dimensional Euclidean space $\mathbb{R}^{3N}$ (with $N$ being the number of \ce{Li+} ions), whose core geometric structures may be essentially described by a lower-dimensional manifold. While the eigenvalue spectrum of the ionic displacement covariance matrix $\bm{C}$ characterizes the relative importance of different collective modes, it does not directly reveal how the orientations of these modes vary across the configurational space. Such a variation could be particularly relevant to the study the motion of the mobile ions in solid electrolytes, in which the transitions from localised vibrational dynamics to free ion conductions may be reflected through the changes in the linearity in this low-dimensional manifold.  

For the investigations of these geometric features, we examine the local tangent spaces associated with neighbouring configurations and quantify their relative orientations through the principal angles between locally estimated collective subspaces. By introducing a normalised measure of the rotation in the  tangent space of the low-dimensional manifold $\mathbb{R}^d$, that is compared to the configurations at controlled separation distances in the original space of $\mathbb{R}^{3N}$, we aim to characterise the temperature-dependent evolution of the local manifold geometry and establish its relationship with the \ce{Li+} ion dynamics in the \ce{Li6PS5Cl}. 

\begin{figure*}[tb]
    \centering
    \includegraphics[width=\linewidth]{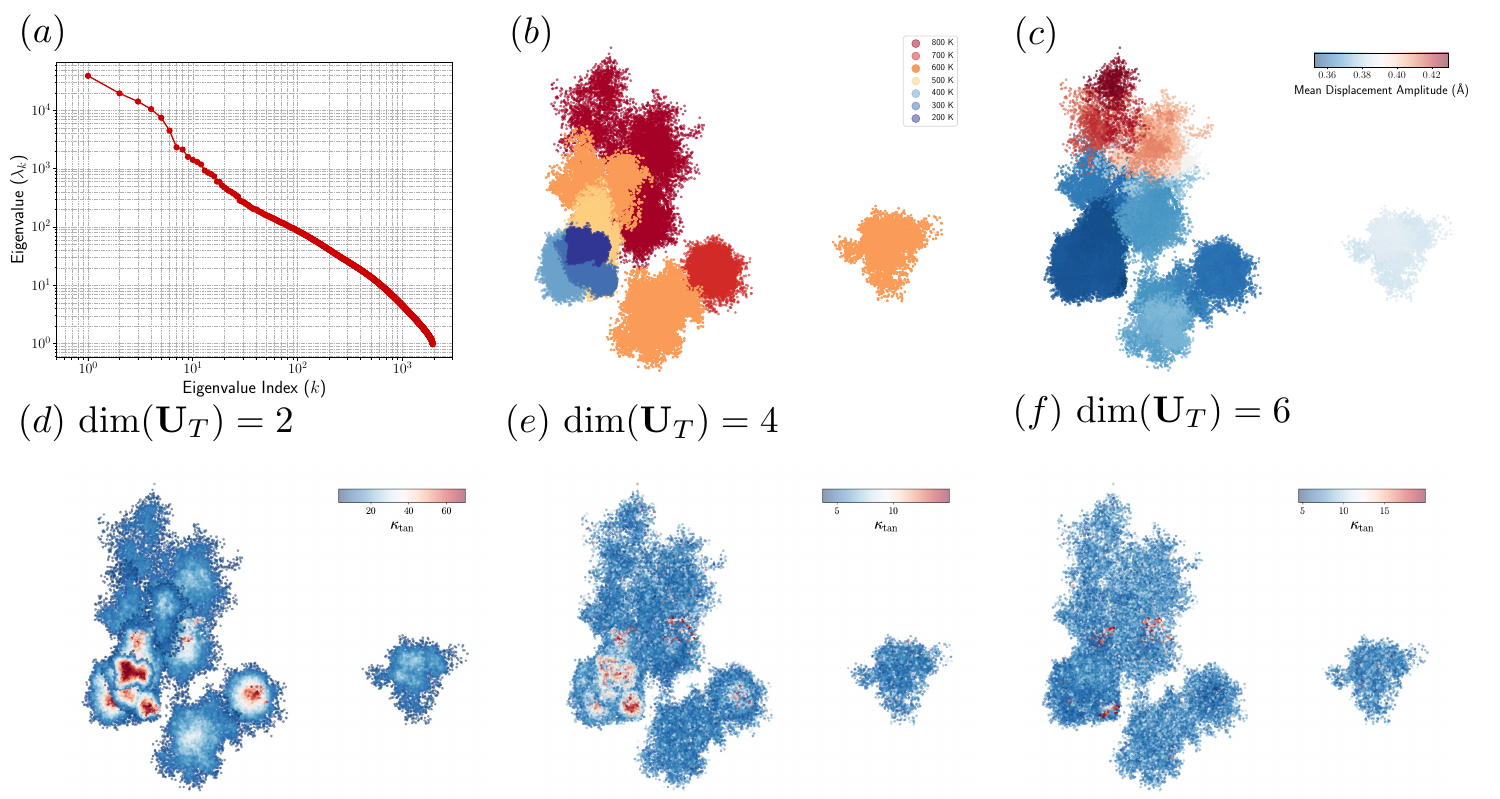}
    \caption{\textbf{Dimensionality reduction analysis of the \ce{Li+} ion displacements.} (a) The eigenvalue spectrum of the multicannonical covariance matrix $\tilde{\bm{C}}_{\mathrm{Li-Li}}$. (b-f) Two dimensional maps of the \ce{Li+} ion displacements produced by projecting the  displacement vector $\Delta\bm{r}_{\mathrm{Li}}$ concatenated for all \ce{Li+} ions in each MD frame onto the first two (principal) eigenvectors of $\tilde{\bm{C}}_{\mathrm{Li-Li}}$, in which each point on the map represents an instantaneous \ce{Li+} ion configuration being sampled by $NVE$ MD. To avoid overcrowding, we only projected MD snapshots that are separated  5 fs apart. Each point on the 2D maps are then colour coded according to: (b) temperatures $T$ at which the MD sampling was performed, (c) the mean amplitude of the displacement that is averaged over the displacements of all the \ce{Li+} ion in the simulation supercell, and (e-f) the local geometric nonlinearity metrics $\kappa_{\mathrm{tan}}$ [\cref{eq:kappa_tan}] that are computed from the projected space $\bm{U}_{T}$ with a dimensionality of 2, 4, and 6, respectively. }
    \label{fig:Li_PCA_maps}
\end{figure*}

To start, we construct $\tilde{\bm{C}}$, dubbed as the \emph{multicanonical} covariance matrix. This is achieved by first concatenating  $NVE$-MD trajectories across multiple temperatures $\tilde{\mathcal{X}}=\mathcal{X}_{T_1}\cup\mathcal{X}_{T_2}\cup\cdots\cup\mathcal{X}_{T_n}$, and computing $\tilde{\bm{C}}$ following the same procedures as outlined in \cref{sec:essential_dynamics} by treating $\tilde{\mathcal{X}}$ as a single MD trajectory. From here, we extract the square sub-block of $\tilde{\bm{C}}_{\mathrm{Li-Li}}$ which characterises how the motions of the \ce{Li+} ions are correlated across all the temperatures simulated. This gives us the corresponding eigenspectrum $\{\tilde{\lambda}_n,\tilde{\bm{v}}_n\}$, with which we can further investigate the geometry of the manifold in which the mobile \ce{Li+} ions vibrate and diffuse. As shown in the eigenvalue spectrum [\cref{fig:Li_PCA_maps}(a)], the first few leading eigenvalues decrease rapidly, similar to what has been observed in \cref{fig:nve_md_essential_dynamics_analysis}. This indicates that the collective dynamics of \ce{Li+} ions lives in a rather low-dimensional geometric manifold, the dimensionality of which can be quantitatively estimated from the partition ratio as $d=(\sum_i \tilde{\lambda}_i)^2/\sum_i \tilde{\lambda}_i^2\approx 9$. 

Subsequently, a two-dimensional embedding $\mathbb{R}^{3N}\to\mathbb{R}^2$ of the \ce{Li+} ion configurations can be constructed by projecting their displacement vectors at each time step [\cref{eq:instant_dispalcement}] onto the first two leading (principal) eigenvectors $\{\tilde{\lambda}_{1},\tilde{\lambda}_{2}\}$ of $\tilde{\bm{C}}_{\mathrm{Li-Li}}$. \cref{fig:Li_PCA_maps}(b) and (c) show this projected map with each point coloured coded according to the sampling temperature $T$, and the mean displacement amplitude of the \ce{Li+} ions $\Vert\Delta \bm{r}(t)\Vert_2/3N$, respectively.  The dominant effect of temperatures on the sampled configurational regions for the \ce{Li+} ions become immediately apparent. At low temperatures (200-400 K), the map shows a strong localisation in the projected configurations along the two principle components, with predominantly low displacement amplitudes. This is consistent with what physically happens in this temperature regime, whereby the \ce{Li+} ion movements can be  primarily  characterised as localised ionic vibrations around their equilibrium and/or metastable lattice positions. At 500 K, the projected configurations into the upper left region with no noticeable enhancement in the displacement amplitudes, indicating that the localised \ce{Li+} vibrations are becoming directionally more diverse. Combining with the previous discussions around \cref{fig:nve_md_essential_dynamics_analysis}(middle), these configurations clearly constitute a \emph{bridging} state, showing notable overlap with projected clusters for the 600 K-configurations, and understandably with more complex geometric structures that cannot be fully captured by just the first two principal components.

The projections for the 600 K-data is particularly interesting, with three isolated clusters, two of which showing little or no overlap with the projected data sampled at the other temperatures. In particular, the rightmost cluster is predominantly occupied with \ce{Li+} ions of mid-ranged displacement amplitudes ($\sim$0.98-0.4 $\mbox{\AA}$), which constitute a set of distinct configurations representing a unique collective displacement patterns that are only observable at 600 K. On the other hand, the middle 600 K-cluster can be considered as representing a metastable transition state, which is the only cluster that overlaps with the projected 700 K-cluster. Similar to the 500 K configurations,  the 700 K configurations are also believed to possess complex geometric structures that may not be fully captured by the first two principal components [\cref{fig:nve_md_essential_dynamics_analysis}(middle)]. Lastly, at 800 K, the cluster extends to the top-left region with the most diverse range of displacement amplitudes.

Overall, by looking at the dominant cluster on the left of the projected map, we can see a significant overlap among clusters across a wide range of sampled temperatures, meaning that many collective motions of the \ce{Li+} ions are accessible across all the temperatures. In another word, the system shares a common configuration space across the temperature range of 200-800 K. As the temperature increases, the projected clusters expands along one particular direction, rather than isotropically. This can be considered as a signature of collective, directional diffusion of \ce{Li+} ions. Lastly, by comparing \cref{fig:Li_PCA_maps}(b) and (c), it is particularly important to note that the displacement amplitude does not correlate simply with the distances to the low temperature clusters. A key reason behind this is that the directional information is lost when taking the norm of the ionic displacements.

The above analyses have provided us with insights on the global geometric signatures of the collective \ce{Li+} ion dynamics. To obtain further insights on the local topologies of the low-dimensional project manifolds, we decorate each point on the projected map with its corresponding local nonlinearity metric $\kappa_{\mathrm{tan}}$ of the $d$-dimensional manifold. Specifically, for a given configuration $\bm{r}_a$ (which, in this case, is represented by a high-dimensional vector of the \ce{Li+} ion displacements, given in \cref{eq:instant_dispalcement}), we first seek for a set of its neighbouring configurations $\{\bm{r}_{b}\}$ with $\Vert\bm{r}_{a}-\bm{r}_{b}\Vert\leq0.8$ $\mbox{\AA}$ with \texttt{scikit-learn}.\cite{scikit-learn} In particular, with restrict $\bm{r}_{a},\bm{r}_{b}\in\mathcal{X}_T$ to be configurations found within the $NVE$ trajectory sampled in the same temperature. Typically, they tend to be a series of consecutive frames in the trajectory. Then we project all configurations onto the lower dimensional manifold, such that in the projected space, we have $\bm{U}_{T}(\bm{r})=(u_1,u_2,\ldots,u_d)$, with $d=\mathrm{dim}(\bm{U}_{T})$, which corresponds to the first $d$ principal components of $\tilde{\bm{C}}_{\mathrm{Li-Li}}$ that are used to construct the projection. In this case, for any two nearby configurations $\bm{r}_{a}$ and $\bm{r}_b$, we can compute the principal angles $\boldsymbol{\theta}_{ab}$ between them in their corresponding tangent space from the singular values $\boldsymbol{\sigma}$ of $\bm{U}_{T}(\bm{r}_{a})^\top\bm{U}_{T}(\bm{r}_{b})$ through $\boldsymbol{\sigma}=\cos\boldsymbol{\theta}_{ab}$. Effectively, $\boldsymbol{\theta}_{ab}$ measures how much the tangent space rotates between these two nearby configurations, which leads to the following measure of local nonlinearity of the projected space:
\begin{equation}
\label{eq:kappa_tan}
    \kappa_{\mathrm{tan}}(\bm{r}_a)=\left\langle \frac{\Vert \sin\boldsymbol{\theta}_{ab}\Vert_{2}}{\Vert\bm{r}_{a}-\bm{r}_{b}\Vert_{2}} \right\rangle_{b\in\mathcal{N}(a)}.
\end{equation}

\cref{fig:Li_PCA_maps}(d) shows the projected map with each point being colour-coded by $\kappa_{\mathrm{tan}}$ with $d=2$, that corresponds to the same dimensionality of the map. Generally, we see that the local non-linearity in the low-dimensional manifold around a low-temperature \ce{Li+} ion configuration being generally and significantly larger than those configurations sampled at higher temperature. This again reflects the localised vibrational behaviours of the \ce{Li+} ions, whereby the (coupled) atomic vibrations around local energy basins typically follows an isoenergetic pathway that is curvilinear.\cite{buvcko2005geometry} In particular, the projected 700 K configurations also show significant local nonlinearity, further reflecting the complex geometric structures that underpins the collective movements of \ce{Li+} ions at this temperature. 
In \cref{fig:Li_PCA_maps}(e,f), we further examined how $d$ and $\kappa_{\mathrm{tan}}$ correlate with each other, which it is clear that, $\kappa_{\mathrm{tan}}$ decreases quickly and systematically as $d$ increases. This reflects that the projection becomes more aligned with the Euclidean metric that characterises the geodistance between the two adjacent \ce{Li+} ion configurations, as in the denominator of \cref{eq:kappa_tan}, which in term also justify $\kappa_{\mathrm{tan}}$ as a meaningful geometric signature for the collective dynamics of the \ce{Li+} ions.

\subsection{Lithium dynamics under the damped framework motions}

\begin{figure*}[tbh]
\centering
\includegraphics[width=\linewidth]{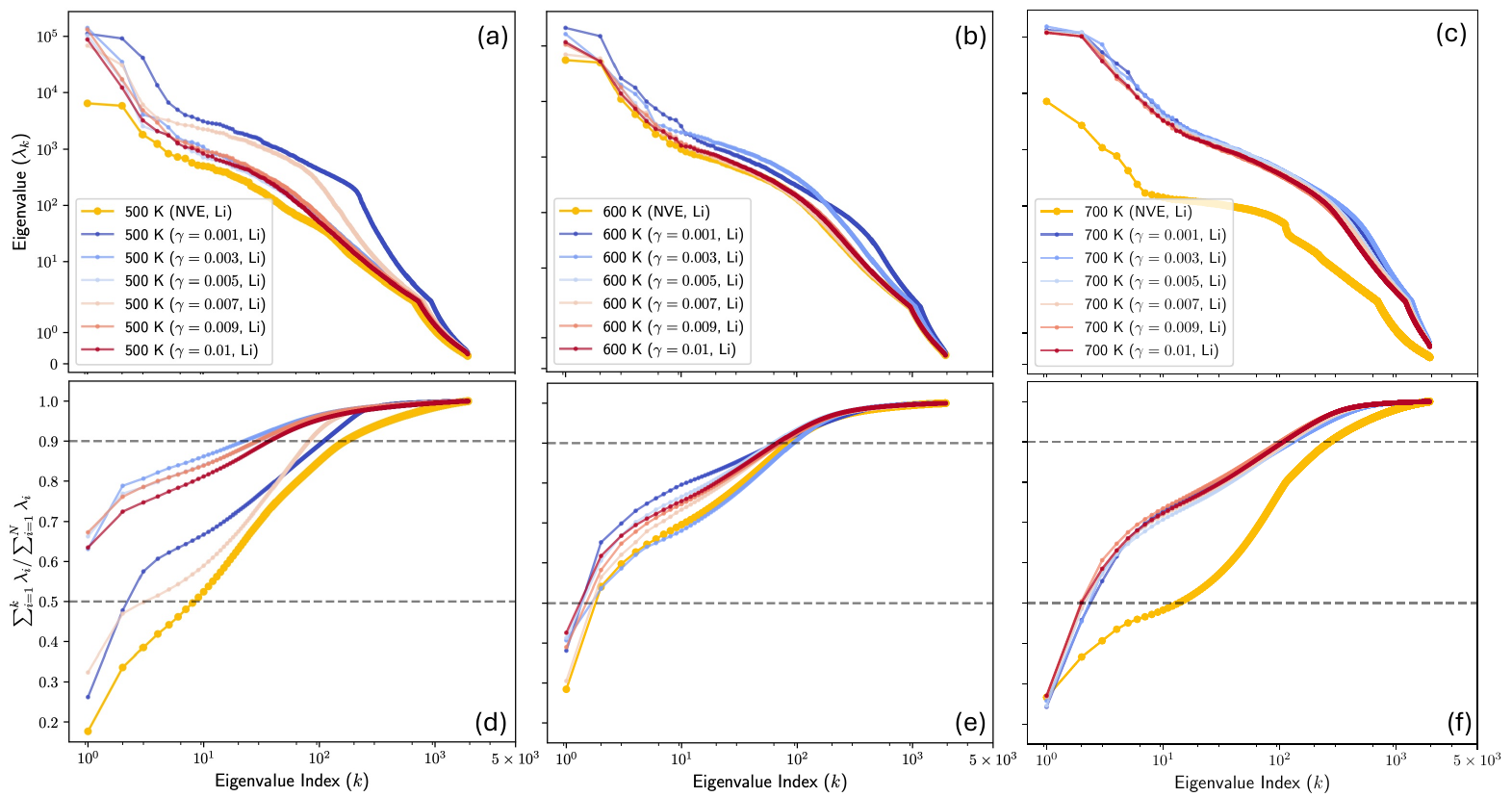}
\caption{\textbf{Friction-parameter-dependent eigenvalue spectrum for essential dynamic analysis based on Langevin MD samplings at 500, 600 and 700 K.} (a-c) Sorted eigenvalues $\lambda_k$  of the covariance matrices characterising the Li-ion-only computed from Langevin MD trajectories with different friction parameters $\gamma$ applied onto the framework ions. (d-f) Participation ratios for the dynamics of Li ions at different friction parameters employed for Langevin dynamics. Yellow curve corresponds to the results obtained from $NVE$ MD simulations (with zero frictional damping).}
\label{fig:langevin_md_essential_dynamics_analysis} 
\end{figure*}

Based on the results from the essential dynamics analysis presented above, we performed additional Langevin dynamic samplings at the selected temperatures of 500, 600 and 700 K, to attain further insights on the interplay between the dynamics of the \ce{Li+}  and framework ions in  \ce{Li6PS5Cl}.  Theoretically, Langevin dynamics\cite{schneider1978molecular} modifies the Newtonian dynamics for ions under the influence of a time-dependent stochastic perturbation $\sigma(t)$ with a frictional term:
\begin{equation*}
    \label{eq:Langevin_dynamic}
    M\ddot{\bm{R}}=-\nabla E(\bm{R})-\gamma M\dot{\bm{R}}+\sigma (t),
\end{equation*}
in which $\gamma$ is the frictional parameter that controls the strength of the interaction between the ion and the heat bath, that effectively damps the dynamics of the ion. 
Here, by  introducing a non-zero $\gamma$ on the \textit{framework ions only} and systematically varying its magnitude when performing the MD sampling, it will provide us with further perspectives  to understand how the dynamics of the \ce{Li+} ions are interacting with those of the framework ions in \ce{Li6PS5Cl} without altering the fundamental dynamics of the system under the $NVE$ ensemble. 
As the eigenvalues are computed for $\bm{C}_{\mathrm{Li-Li}}$ only, the results obtained from the Langevin dynamics with varying $\gamma$ (as compared to those from $NVE$ MD) will further provide us with the key indicators on how strongly the Li ion dynamics are affected by the dynamics of  framework   before and after the material enters the superionic  state. \cref{fig:langevin_md_essential_dynamics_analysis} presents the results obtained from such analysis. 

It can be seen from \cref{fig:langevin_md_essential_dynamics_analysis}(b,e) that, at 600 K, the critical temperature at which liquid-like \ce{Li+} ion diffusion is activated, damping the the dynamics of the framework ions has exerted negligible effects on the \ce{Li+} ion dynamics, regardless of the magnitude of the chosen damping factor $\gamma$. Such a result is clearly expected when the participation ratios from the framework ions are reduced (\cref{fig:nve_essential_dynamics_atom_contributions}).
The resulting effects on the \ce{Li+} ion dynamics upon damping the framework motions at 500 K [\cref{fig:langevin_md_essential_dynamics_analysis}(a,d)] and 700 K [\cref{fig:langevin_md_essential_dynamics_analysis}(c,f)] are clearly different from those observed at 600 K. 

At 500 K, we see that, upon introducing a small damping factor ($\gamma=0.001$), a strong increase (approx. an order of magnitude) in $\{\lambda_{\mathrm{Li-Li}}\}$ across the whole eigenvalue spectrum can be observed. As the damping parameter $\gamma$ was increased, the first eigenvalue $\lambda_1$ remains largely unchanged, and is more than an order of magnitude larger than $\lambda_1$ from the $NVE$ result (yellow curve in \cref{fig:langevin_md_essential_dynamics_analysis}). However, the subsequent eigenvalues $k\geq2$ gradually reduce back to around the same level as the distribution of $\{\lambda_{\mathrm{Li-Li}}\}$ that were obtained from $NVE$ sampling performed at 500 K. \jy{The most likely explanation behind this result, is that, at 500 K, there is only a small portion of low-energy framework vibrations that are acting against the liberation of \ce{Li+} ions to entering a fast conducting state, thus damping these framework dynamics is seemingly promoting the \ce{Li+} conductions at a lower temperature. Such a conducting stage for the \ce{Li+} ions is likely to be facilitated by certain higher energy framework vibrations, which may explain why, upon further increase of $\gamma$, the motions of \ce{Li+} are suppressed again, as reflected in the reduction of $\{\lambda_{\mathrm{Li-Li}}\}$. It also indicates that the amount of thermal energy at 500 K is insufficient for the \ce{Li+} ions to overcome the diffusion barriers imposed by the crystal framework.}

At 700 K, the system is expected to have entered another dynamic regime, at which the framework can be considered as moving with the \ce{Li+} ions concomitantly, where the large amplitude motions of the framework slow down the Li motions. \jy{In the TDEP\cite{hellman2013temperature} framework, we may consider the chemical environment created by the fluctuating framework ions, which are now activated by the sufficient amount of thermal energy, are now acting collectively as some \emph{frictional} forces that are trying to slow down the migration of the \ce{Li+} ions again.}
This can be seen in \cref{fig:langevin_md_essential_dynamics_analysis}(c,f) on why, upon a small damping of the framework dynamics, the \ce{Li+} ions are \emph{liberated} again, leading to the observed increase in $\{\lambda_{\mathrm{Li-Li}}\}$ over the entire range of $k$, and remain largely insensitive to the magnitudes of the damping parameter $\gamma$. 

An interesting question to examine at this point is that to what extent that the conductions of the \ce{Li+} ions are mediated by the \emph{anharmonic} vibrations of the frameworks.\cite{yang2026mace,knoop2020anharmonicity} However, when examining the statistical distributions of the framework ion displacements across the simulated temperatures [\cref{fig:FM_displacement_distribution_analysis}], they all appear to follow the Gaussian distribution nicely, suggesting that the dynamics of the framework do not exhibit notable anharmonicity across the simulated temperatures, thus we shall not proceed further with discussions along this line.

\subsection{Temporal correlations in the ion dynamics}

In order to gain further insights into the temporal correlations in the dynamics of the mobile Li and framework ions in \ce{Li6PS5Cl}, we compute and analyse the lag-time-dependent covariance matrix $\bm{C}(\tau)$ at each temperature $T$ with its matrix elements defined as 
$C_{ij}(\tau)=\left\langle  \sqrt{m_i m_j}\Delta \bm{r}_i(t,\tau)\cdot\Delta \bm{r}_j(t,\tau)\right\rangle_t$, in which $\Delta \bm{r}_i(t,\tau)=\bm{r}(t+\tau)-\bm{r}(t)$ (where the drift of the centre-of-mass with respect to the averaged structure $\langle\bm{r}\rangle_t$ was removed, similar to the analysis performemd above). $\tau$ is the lag-time, and  scanning $\tau$ across a range from $\sim 10^{-1}$ to $\sim10^2$ ps, will enable us to understand how the ionic motions are correlated at the time scales that are relevant to typical phonon vibrations (short $\tau$) as well as long-ranged Li diffusions (large $\tau$). To facilitate the analysis, we first compute separately the matrices that correlate the Li-Li interactions (by restricting both $i$ and $j$ to run through only the indices of atoms in the Li-sublattices) and the Li-framework interactions (by restricting both $i$ and $j$ to run through only the indices of atoms in the Li-sublattices and the framework, respectively). Each matrix element is then properly normalised as
\begin{equation}
    \label{eq:normalised_rho}
    \rho_{ij}(\tau)=\frac{\left\langle\sqrt{m_i m_j}\Delta \bm{r}_i(t,\tau)\cdot\Delta \bm{r}_j(t,\tau)\right\rangle_t}{\sqrt{\langle \sqrt{m_i}|\Delta r_i(t,\tau)^2|\rangle_t \langle \sqrt{m_j}|\Delta r_j(t,\tau)^2|\rangle_t}}
\end{equation}
Physically, the sign of this matrix element provides an indication on whether the motions of two ions are concerted, weakly correlated or anticorrelated, which correspond to the cases of $\rho_{ij}<0$, $\rho_{ij}\sim0$ and $\rho_{ij}>0$, respectively. From here, one can further construct  \emph{global} temporal correlation metrics that characterise the dynamics of all $N$ pairs of ions in the crystal:
\begin{equation}
    \label{eq:global_metric}
    \begin{cases}
        \rho_{\mathrm{net}}(\tau) = \displaystyle{\frac{2}{N(N-1)}\sum_{i<j}\rho_{ij}(\tau)}\\
        \rho_{\mathrm{abs}}(\tau) = \displaystyle{\frac{2}{N(N-1)}\sum_{i<j}\left\vert\rho_{ij}(\tau)\right\vert}.
    \end{cases}
\end{equation}
Under such definitions, $\rho_{\mathrm{net}}$ indicates whether the overall motions of the ions appear predominantly cooperative or anti-cooperative, whereas $\rho_{\mathrm{abs}}$ measures how strongly are the motions of the ions correlated with each other on average.
        
\begin{figure*}[tbh]
    \centering
    \includegraphics[width=\linewidth]{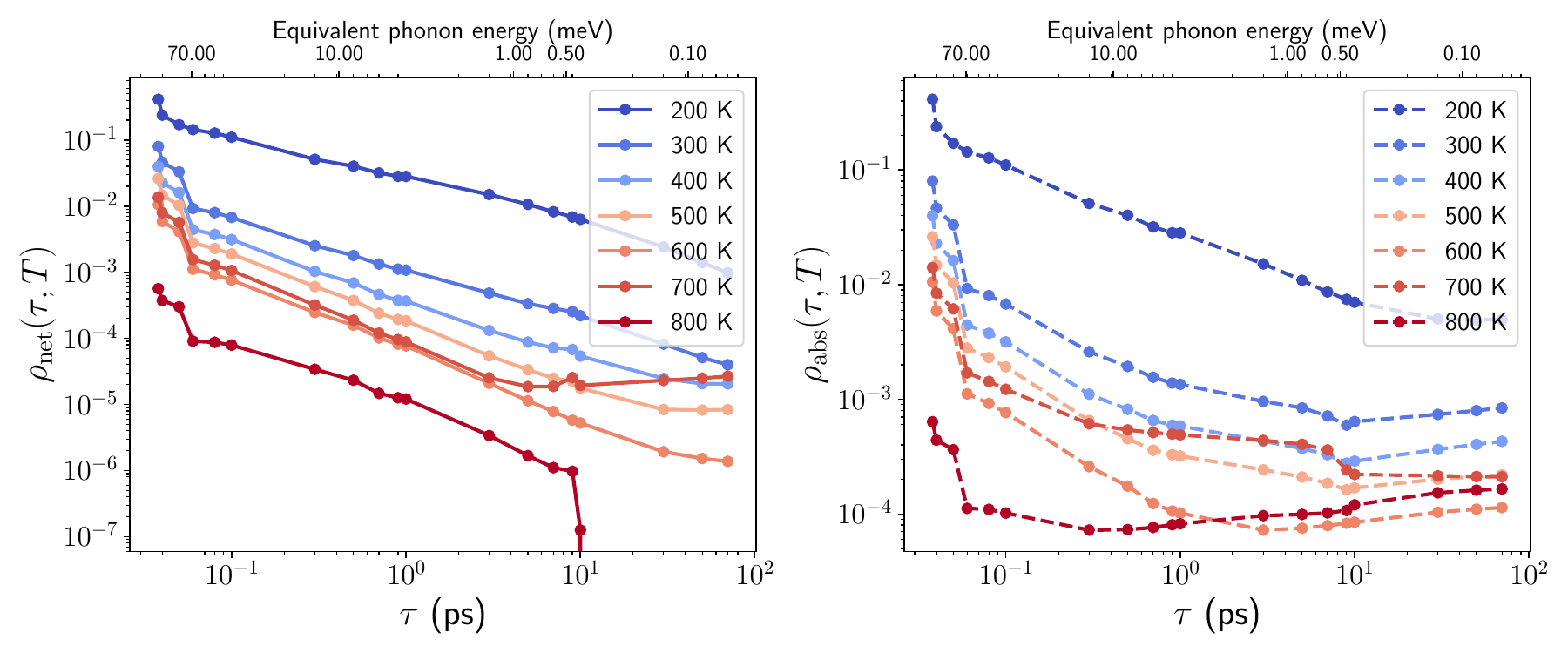}
    \caption{\textbf{Correlations of lithium ion dynamics in \ce{Li6PS5Cl}.} Lag-time and temperature dependent net [$P_{net}(\tau,T)$, left] and absolute [$P_{abs}(\tau,T)$, right] correlation scores characterising the degrees of correlated motions for the \ce{Li+} ions in \ce{Li6PS5Cl}. The top $x$-axes indicate the  equivalent phonon energies with respect to the corresponding lag-time $\tau$.}
\label{fig:Li_displacement_correlation_vs_tau}
\end{figure*}

\begin{figure*}[tbh]
    \centering
    \includegraphics[width=\linewidth]{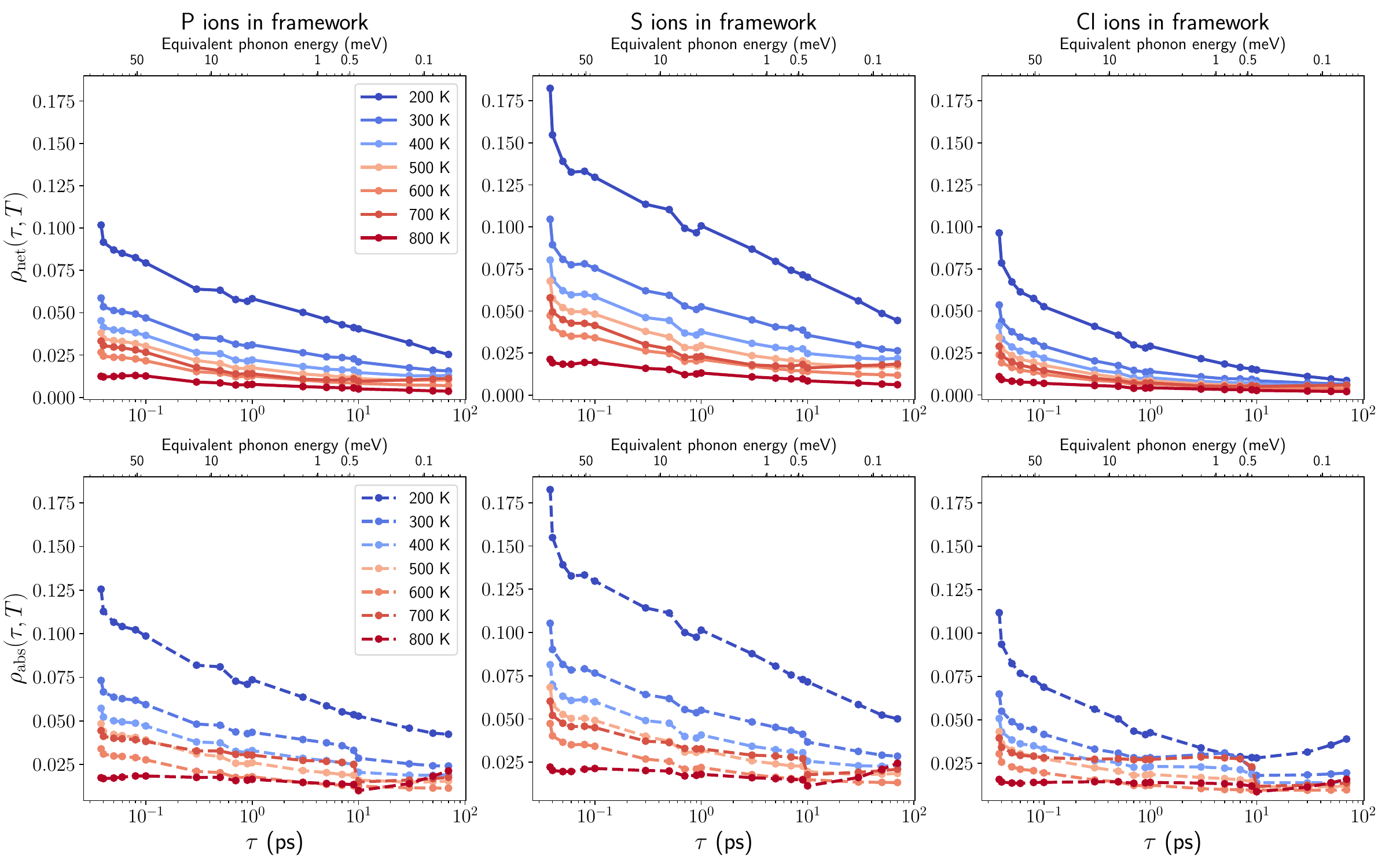}
    \caption{\textbf{Correlations of the lithium and framework dynamics in \ce{Li6PS5Cl}.} Lag-time and temperature dependent net [$P_{net}(\tau,T)$, top] and absolute [$P_{abs}(\tau,T)$, bottom] correlation scores characterising the degrees of correlated motions among  the \ce{Li+} and framework ions in \ce{Li6PS5Cl}.}
    \label{fig:Li_framework_displacement_correlation_vs_tau}
\end{figure*}

\cref{fig:Li_displacement_correlation_vs_tau}  presents the metrics measuring the correlations between the motions of pairs of \ce{Li+} ions as a function of the lag-time $\tau$ at different temperatures under which the $NVE$ MD samplings were performed. Generally, the trends of $\rho_{\mathrm{net}}$ and $\rho_{\mathrm{abs}}$ largely follow each other, indicating that the  motions of \ce{Li+} ions are cooperative in the crystal. It is clear that, at $\tau\sim0.1$ ps, the motion of the \ce{Li+} ions are about five times more correlated at 200 K than that at 300 K, reflecting that, at low temperature, the \ce{Li+} ions are predominantly confined within the local potential energy basins, where their vibrational modes strongly govern their thermal motions. This is because the latter is fundamentally determined by the Hessian matrix that encodes only the two-body correlations in the lattice dynamics. Consequently, we can also see the strongest correlations between the vibrations of the \ce{Li+} and framework ions at 200 K across all $\tau$ [\cref{fig:Li_framework_displacement_correlation_vs_tau}], and such correlations are  approximately halved in magnitudes as the temperature is raised to 300 K. As $\rho_{\mathrm{net}}$ decreases monotonically as a function of $\tau$ across all temperatures, it is evident that any correlations in the motions between pairs of \ce{Li+} ions in the crystal are predominantly mediated by the fast phonon vibrations. Nevertheless, it can be seen from \cref{fig:Li_displacement_correlation_vs_tau}, that at large $\tau$, the values of $\rho_\mathrm{abs}$ are slightly higher than  $\rho_\mathrm{net}$, particularly at the low temperature, suggesting a small degree of anti-cooperative \ce{Li+} ion dynamics at long timescales.

The reductions in the degrees of correlations in the motions of the \ce{Li+} ions become less significant with respect to the rising temperature above 300 K, and such correlations can be considered to be almost negligible across the full range of $\tau$ at 600 K and above. This can be considered as a signature of the liquid-like \ce{Li+} conduction behaviours in \ce{Li6PS5Cl} superionic conductor, whereby the thermal energy quickly destroy the correlated motions of the mobile ions. Nevertheless, the result shown here does not necessarily mean that correlated \ce{Li+} ion hopping will not occur transiently and locally throughout the crystal at the elevated temperatures, but any contributions from such events tend to diminish quickly upon statistical averaging.

In \cref{fig:Li_framework_displacement_correlation_vs_tau}, we further analyse the temporal correlations between the motions of the \ce{Li+} and framework ions. Compare to \cref{fig:Li_displacement_correlation_vs_tau}, it can be seen that the correlations between the \ce{Li+} and framework ions are stronger than those between pairs of \ce{Li+} ions. In particular, the correlations between \ce{Li+} and S$^{2-}$ ions are the strongest, as the S$^{2-}$ ions are directly coordinating with \ce{Li+}  in the crystal structure (\cref{fig:structure}). It can be seen from the top panel of \cref{fig:Li_framework_displacement_correlation_vs_tau} that, the metric $\rho_\mathrm{net}$ generally decreases against increasing $\tau$, indicating that the fast phonon-like vibrations of the framework still plays a dominant role in the coupling with the motions of the \ce{Li+} ions. We also see that $\rho_\mathrm{net}$ between \ce{Li+} and all three framework ions  decrease as temperature rises, as for the correlations between pairs of \ce{Li+} ions. As for the correlated motions among this \ce{Li+} ions, this does not necessarily indicate the  diminishing of phonon-mediated coupling between \ce{Li+} and framework motions, but suggesting the at higher temperatures, the ionic motions are more prone to be randomised by the thermal excitations, making them making them appear to be less directionally correlated.

\begin{figure*}[htb]
    \centering
    \includegraphics[width=\linewidth]{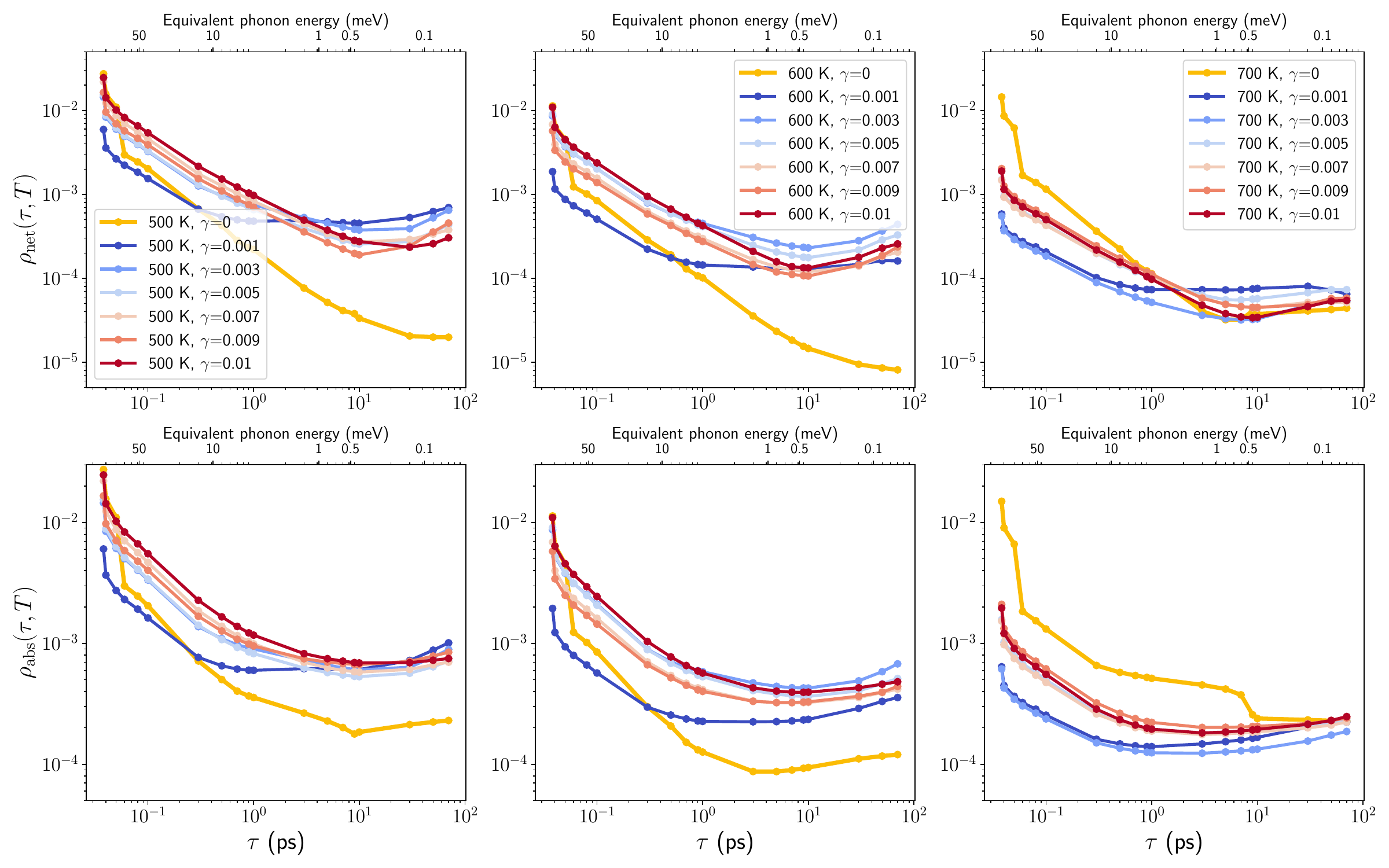}
    \caption{\textbf{Correlation of lithium ion dynamics in \ce{Li6PS5Cl} under the damped framework dynamics.} Lag-time and damping parameter dependent net [$P_{net}(\tau,T)$, top] and absolute [$P_{abs}(\tau,T)$, bottom] correlation scores characterising the degrees of correlated  motions for the \ce{Li+} ions in \ce{Li6PS5Cl}.}
\label{fig:Li_displacement_correlation_vs_tau_langevin}
\end{figure*}

Finally, we repeat the analysis on the temporal correlations in the \ce{Li+} ion dynamics on the trajectories of Langevin dynamics with different damping strengths on the framework ions. The results shown in \cref{fig:Li_displacement_correlation_vs_tau_langevin} reveal that:  At 500 and 600 K, increasing $\gamma$ generally enhances the short-timescale correlations between the motions of pairs of \ce{Li+} ions, although such an increase is not strictly monotonic with respect to $\gamma$ (which is similar to the result as presented in \cref{fig:langevin_md_essential_dynamics_analysis}). On the other hand, at 700 K, the behaviours change fundamentally, in which the zero-damping trajectory exhibits the strongest short-time correlations, whereas large $\gamma$ suppresses them (which is opposite from the result  presented in \cref{fig:langevin_md_essential_dynamics_analysis}). Across all three temperatures, it can also be seen that the correlation metrics $\rho$ are generally insensitive to the damping parameters $\gamma$ within the intermediate range of $\tau$, but notable uprise in $\rho_{\mathrm{abs}}$ are noticeable at long $\tau$, particularly at 700 and 800 K. In conjunction with the results presented in \cref{fig:langevin_md_essential_dynamics_analysis}, we believe that: Below 700 K, constraining the framework dynamics does tend to \emph{liberate} the large amplitude motions of the \ce{Li+} ions in the crystal, but even at the onset of the critical temperature in which the system enters the superionic conducting stage, there is still some characters of the phonon-like motion being retained by the mobile \ce{Li+} ions (albeit quantitatively halved at 600 K with respect to that of 500 K as revealed in \cref{fig:Li_displacement_correlation_vs_tau_langevin}). At 700 K, however,  constraining the framework dynamics  fully \emph{liberate} the \ce{Li+} ions in which the motions of the conducting \ce{Li+} ions become directionally uncorrelated in a statistical measure, further reflecting the `liquid-like' fast-ion conducting state at this temperature.

\section{Conclusions and Future Outlooks}

Superionic conductors with liquid-like conducting behaviours constitute an interesting material system with many intriguing dynamic properties that are yet to be explore at the intersection between an ordered solid and a liquid lacking long-ranged orderings. Recent advances in foundational materials models have opened up new opportunities for us to deploy large-scale and long-time MD simulations, with additional physical constraints, to systematically study the dynamics of the mobile \ce{Li+} ions, and how they are interacting with the dynamics of the crystal frameworks in superionic conductors, which is demonstrated in the present work, focusing on the argyrodite solid electrolyte \ce{Li6PS5Cl}.  In particular, we capture the temporal and spatial correlations in the ionic movements through constructing the (time-dependent) displacement correlation matrix, following the conceptual framework of essential dynamic analysis, that provides a single mathematical construction to examine the ion dynamics endowing both the localised phonon-like and long-ranged diffusive regimes.

Our key findings are of several folds. On the foundation energy models, our study presents another key scenario in which further fine-tuning the model is necessary to correctly capture the lattice and ion dynamics in superionic conductors,\cite{fragapane2026li} as the mobile ions traverse a must wider configurational space, leading to a broad distribution in the local environments that are generally not present in the training data that are primary collated from the local geometry optimisation trajectories. In terms of the material \ce{Li6PS5Cl} itself, our analysis finds that: (a) At the the low temperature (200-400 K) regime, the \ce{Li+} ion motions are predominantly of phonon-like with strong nonlinearity in the low-dimensional manifold that characterises their dynamics, which is of the completely opposite case at the higher temperature ($\sim$ 800 K), where the \ce{Li+} ions undergo large amplitude diffusive motions that give rise to a trajectory with higher local linearity even when it is projected to a two-dimensional manifold. (b)  Within the temperature range of 500-700 K, crossing the onset of the liquid-like conducting state transition temperature, the ion dynamics and their corresponding geometric characteristics  become more complex, signifying the \ce{Li+} ions trying to overcome the competing displacive framework dynamics to become freely conducting. This is further reflected in the detailed analysis of the Langevin dynamics and the temporal correlations in the \ce{Li+} ion displacements. 

We note that many previous studies on \ce{Li6PS5Cl} and related superionic conductors\cite{hanghofer2019fast,jun2024diffusion,jun2024nonexistence} played a strong focus on the coupling between the paddle-wheel or soft-cradle-craft motions of the PS$_4^{3+}$ motifs with those of the \ce{Li+} ions. While in our analysis, it is possible to further group the motions of P and S together to construct collective motif-based coordinates and examine their coupling with the \ce{Li+} ions in the Euclidean space, this approach is not pursued in the present work, as we would like to keep the methodology as general as possible by being not constrained to a material-specific coordinate system. 

Moving forward, we expect the continuous development in foundation models for materials will further enhance our capabilities in studying the dynamic responses of more complex materials, with the possible inclusions of the effects from external fields.\cite{batatia2026mace} Extending the current two-body correlation matrices to tensors that encoding higher-body-order effects, particularly overcoming the computer memory bottlenecks in handling these tensors (for example, through a cut-off-radius-based scheme), would also be interesting to study systems in which the framework itself possesses strong anharmonic vibration characteristics.\cite{gupta2021fast,ahammed2024configurational} Lastly, from a pure theoretical perspective, exploring mathematical approaches, such as those based on differential geometry, to characterise the  diffusion pathways of mobile ions in solid electrolytes, thus facilitating the further development of our understanding on how ionic diffusions are affected by the materials' structures, would be an interesting future research direction.   

\section*{Data Availability}

The computer code and fine-tuned MACE model associated with this work can be found in the GitHub Repository: \url{https://github.com/yangjackie/diffuser}.

\section*{Acknowledgements}

JY acknowledges the support from the Australian Research Council (ARC) Discovery Project of DP250100778. AL acknowledges the support of the National Natural Science Foundation of China under grant No. 12264017. SL acknowledges the support of the ARC Discovery Projects of DP250104633 and Laureate Fellowship of FL250100143. Computations in this work are conducted on the Gadi supercomputer at NCI Australia that is supported by the UNSW Merit Allocation Scheme under the project \texttt{dy3}. 

\bibliography{ref.bib}

\newpage
 
\setcounter{section}{0}
\setcounter{figure}{0}

\renewcommand{\thesection}{S\arabic{section}}
\renewcommand{\thesubsection}{S\arabic{section}.\arabic{subsection}}

\renewcommand{\thefigure}{S\arabic{figure}}

\section*{Appendix}

\begin{figure*}[htb]
    \centering
    \includegraphics[width=0.7\linewidth]{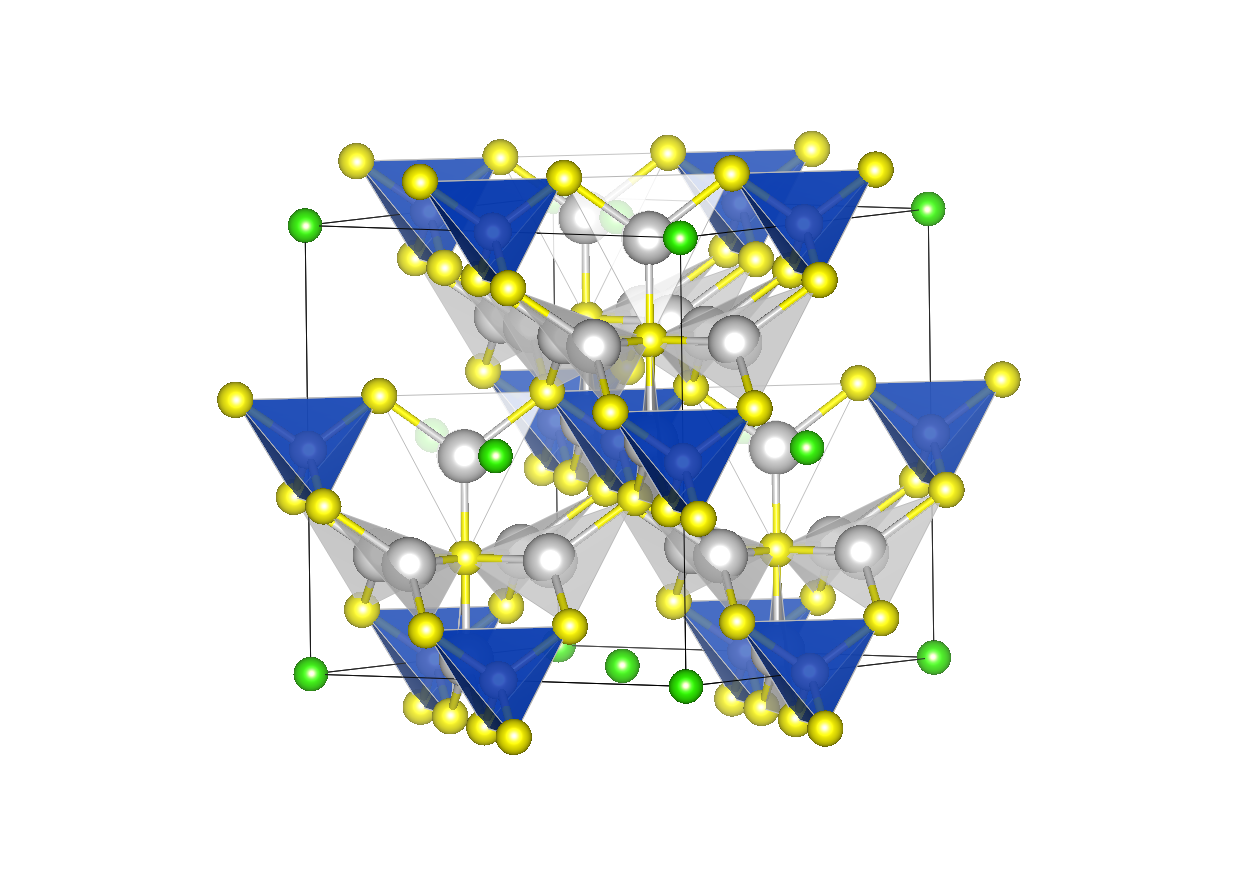}
    \caption{\textbf{Crystal structure of cubic \ce{Li6PS5Cl}.} Atoms are color coded as: silver (Li), blue (P), yellow (S) and green (Cl).}
    \label{fig:structure}
\end{figure*}

\begin{figure*}
    \centering
    \includegraphics[width=0.6\linewidth]{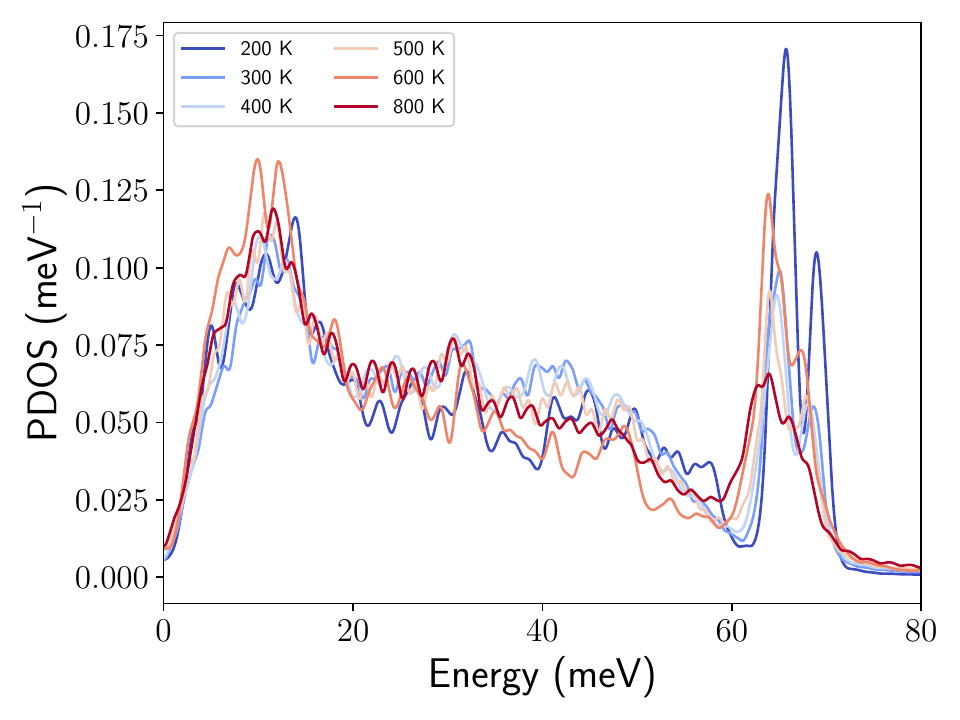}
    \caption{\textbf{Phonon density-of-states for \ce{Li6PS5Cl}.} Total phonon density-of-states at different temperatures computed from the Fourier transform of the velocity autocorrelation functions based on $NVE$-MD samplings performed with the fine-tuned MACE foundation model.}
    \label{fig:placeholder}
\end{figure*}

\begin{figure*}
    \centering
    \includegraphics[width=\linewidth]{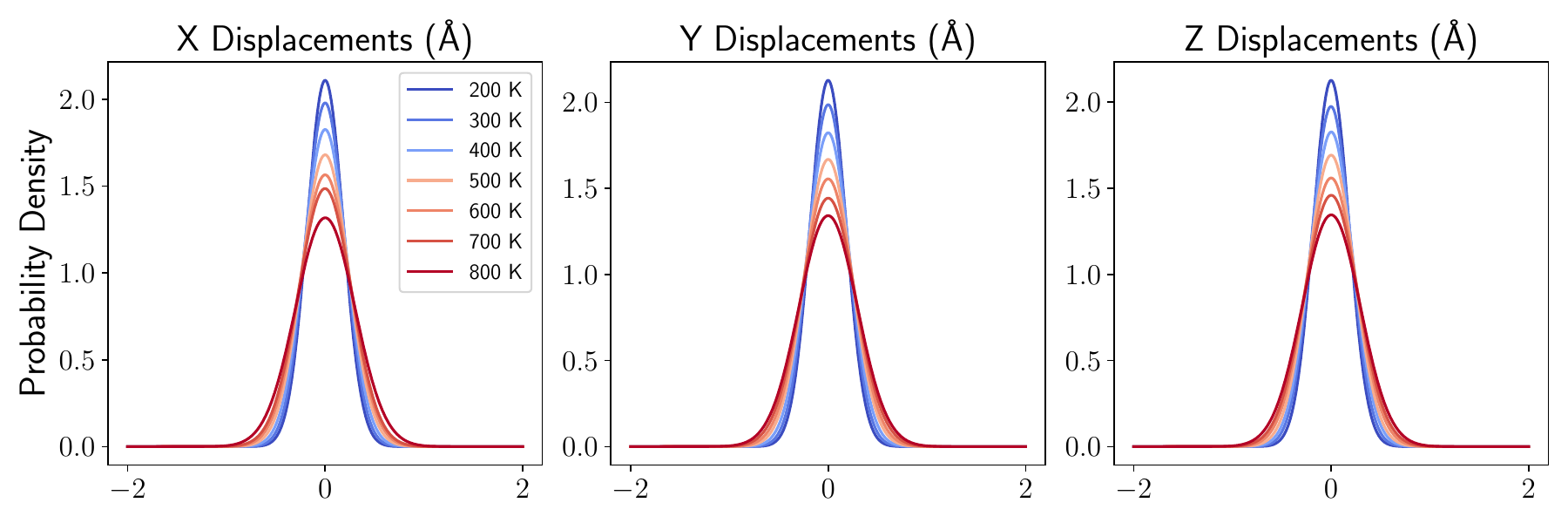}
    \caption{\textbf{Vibrational characteristics of the framework ions.} Statistical distributions of the displacements of all framework ions along the three Cartesian directions as sampled in $NVE$ MD simulations at different temperatures. The apparent Gaussian distributions in the ionic displacements indicate that the framework vibrations are predominantly harmonic,\cite{knoop2020anharmonicity} and are progressively widened as the temperature increases, reflecting the increasing vibration amplitudes at higher temperatures.}
\label{fig:FM_displacement_distribution_analysis}
\end{figure*}

\end{document}